\documentclass[12pt, a4paper]{article}
\usepackage[utf8]{inputenc}
\usepackage{physics} 
\usepackage{amsmath}
\usepackage{amsfonts}
\usepackage{amssymb}
\usepackage{graphicx}
\usepackage{authblk}
\usepackage{cite}

\newcommand {\ve}{\varepsilon}

\newcommand {\cG}{\cal G}
\newcommand {\cD}{\cal D}
\newcommand {\bg}{\bar \gamma}
\newcommand {\bp}{\bar \psi}

\usepackage[left = 1.5cm, right = 1.5cm, top = 1cm, bottom = 2cm]{geometry}
\begin{document}

\begin{center}
{}

{\large {\bf Constraining parameters of spinor field dark energy: An alternative to $\Lambda CDM$ model under the spherically symmetric FLRW space-time}}
\vskip .5cm
\textbf{Bijan Saha$^{a,b}$ Mahendra Goray$^{c,d}$}
\end{center}
\vskip0.1cm
$^a$Laboratory of Information Technologies, Joint Institute for Nuclear Research,
141980 Dubna, Moscow region, Russia\\
$^b$Peoples' Friendship University of Russia (RUDN University), 6 Miklukho-Maklaya Street, Moscow, Russian Federation\\
$^c$Department of Mathematics, Central University of Odisha, Koraput-763004, Odisha, India\\
$^d$Department of Physics, S.P. College, Sido Kanhu Murmu University, Dumka-814101, Jharkhand, India\\

\textbf{E-mail:} bijan@jinr.ru \textbf{(B.Saha)}, goraymahendra92@gmail.com \textbf{(M.Goray)}\\

\abstract{This study constrains the cosmological parameters within the scope of a spherically symmetric FLRW cosmological model, the role of a nonlinear spinor field in the universe's evolution. To test this approach, we incorporate the recent Cosmic chronometers, Supernova, and Sloan Digital Sky Survey data. It is found that if spherical coordinates give the FLRW model, the energy-momentum tensor (EMT) of the spinor field possesses nontrivial non-diagonal components. These non-diagonal components of EMT neither depend on the spinor field nonlinearity nor the value of the parameter $k$ defining the type of curvature of the FLRW model. In this context, we construct a dark energy model and perform an MCMC simulation to obtain the best-fit values of the parameters. The results are well comparable to the present Hubble parameter and deceleration parameter, indicating the accelerated expansion of the universe.
}.

\section{Introduction}
In present-day cosmology, numerous theoretical models have been studied to account for the universe's accelerated expansion \cite{andriot2023accelerated, Padmanabhan2002, B7}. Observational evidence from the Type Ia supernova (SNIa) \cite{Riess1998, Perlmutter1999}, Cosmic Microwave Background (CMB)\cite{Planck2013, Planck2018}, Baryon Acoustic Oscillations (BAO), Surveys for growth of Large-Scale Structure (LSS), etc.\cite{Anderson2014, Alam2017, eBOSS2021, DESI2024, DESI2023}, has confirmed the acceleration of our universe.  Even two and a half decades after discovering the accelerated universe, the reason behind it is still a mystery. However, it has been considered that an exotic component with negative pressure is responsible for the acceleration. Namely, it is called dark energy (DE); presently, it contains around two-thirds of the energy density of the universe \cite{Riess1998, Perlmutter1999, Planck2018, DESI2024}.  

Spinor fields are gaining attention as powerful tools for modeling various cosmic phenomena, such as dark energy and the accelerated expansion of the universe. Their versatility in representing different forms of matter makes them valuable for exploring the universe’s evolution across different epochs
\cite{kremer1,Saha2006PRD,Saha2006,Saha2009ECAA,FabIJTP,ELKO,FabJMP,PopPRD}.
It has been observed that the spinor field is highly sensitive to the geometry of spacetime. Depending on the specific form of the metric, the spinor field can exhibit various nontrivial, non-diagonal components in its energy-momentum tensor. Consequently, the spinor field imposes distinct constraints on both the structure of spacetime and the properties of the field itself \cite{2018EChAYa49-146-212}. 
Recently, spinor fields have been employed in astrophysical research to explore whether their distinctive properties can offer new perspectives on phenomena like black holes and wormholes. These investigations have been conducted within the framework of spherically symmetric \cite{Saha2018, BRS2020} and cylindrically symmetric spacetime \cite{Spincyl, Spinepjp22}.

Since that the present-day universe is remarkably isotropic, and that nontrivial non-diagonal components of the spinor field impose strong constraints on its behavior, we also examined the role of spinor fields within the Friedmann–Lemaître–Robertson–Walker (FLRW) model. In those studies, spacetime was represented using Cartesian coordinates. To investigate how coordinate transformations affect the spinor field, we previously conducted additional research on this aspect\cite{zhenya,arsenyi}. 

In this work, we will further analyze the cosmological data sets under the framework of the spherically symmetric FLRW model filled with a spinor field. This isotropic and homogeneous cosmological FLRW model generates the Chaplygin gas with an equation of state (EoS) in the form of pressure $p = W \ve - A/\ve^\alpha, \label{mchap}$ with $A > 0$ and $0 \le \alpha \le 1$ \cite{bsaha2025}. This model is labeled as Modified Chaplygin Gas (MCG), which could be a prominent candidate for the dark energy in the range of $-1<W<A$, that beautifully incorporates with the late-time universe. 

We parametrize the Hubble parameter $H(z)$, EoS, and all other cosmological parameters directly related to the spinor field DE model. To constrain the model parameter, we used the Markov Chain Monte Carlo (MCMC) statistical method in a Python 3.0 environment. The MCMC technique has been performed with the help of SNIa, Cosmic Chronometers (CC), and BAO  data sets. More specifically, we combined supernova(binned) data, Observational Hubble Data (OHD), and Sloan Digital Sky Survey (SDSS) data sets, besides we simulated the full Pantheon data set only. Also, we compare our results with the $\Lambda$ Cold Dark Matter ($\Lambda CDM$) model for the evolution of $H(z)$.

Our work is organized in this paper as follows. Mathematical background of the MCG model in Section 2. Section 3 is devoted to dealing with the observational data and the MCMC method to constrain the parameters. Next section, analyze the numerical results. The conclusions are drawn in Section 5. 

\section{Theoretical background}
\subsection{Basic equations}
The action we choose in the form
\begin{eqnarray}
{\cal S} = \int \sqrt{-g} \left[\frac{R}{2 \kappa} + L_{\rm sp}
\right] d \Omega, \label{action}
\end{eqnarray}
where $\kappa = 8 \pi G$  is Einstein's gravitational constant, $R$
is the scalar curvature and $L_{\rm sp} $ is the spinor field
Lagrangian given by \cite{SahaPRD2001}
\begin{eqnarray}
L_{\rm sp} = \frac{\imath}{2} \biggl[\bp \gamma^{\mu} \nabla_{\mu}
\psi- \nabla_{\mu} \bar \psi \gamma^{\mu} \psi \biggr] - m \bp \psi
- \lambda F(K). \label{lspin}
\end{eqnarray}

To maintain the Lorentz invariance of the spinor field equations, the
nonlinear term $F(K)$ in \eqref{lspin} is constructed as some
arbitrary functions of invariants generated from the real bilinear
forms. On account of Fierz equality in \eqref{lspin} we set $K =
K(I, J) = b_1 I + b_2 J,$ where $b_1$ and $b_2$ takes the value $0$
or $1$, which leads to the following expressions for $ K =
\{I,\,J,\,I+J,\,I-J\}$. Here $I = S^2$ and $J = P^2$ are the
invariants of bilinear spinor forms with $ S = \bp \psi$ and $P =
\imath \bp \bg^5 \psi$ being the scalar and pseudo-scalar,
respectively. In~\eqref{lspin} $\lambda$ is the
self-coupling~constant. Note that $\lambda$ can be both positive and negative, while 
$\lambda = 0$ leads to the linear case. Here $m$ is the spinor mass.

The covariant derivatives of a spinor field take the form
\cite{SahaPRD2001}

\begin{align}
\nabla_\mu \psi = \partial_\mu \psi - \Omega_\mu \psi, \quad
\nabla_\mu \bp =
\partial_\mu \bp + \bp \Omega_\mu, \label{CVD}
\end{align}

where $\Omega_\mu$ is the spinor affine connection, defined as
\cite{SahaPRD2001}

\begin{equation}
\Omega_\mu = \frac{1}{4} g_{\rho \sigma} \left(\partial_\mu
e^{(b)}_\tau e^\rho_{(b)} - \Gamma_{\mu
\tau}^\rho\right)\gamma^\sigma \gamma^\tau. \label{SPAC}
\end{equation}

In \eqref{SPAC} $\Gamma^{\beta}_{\mu\alpha}$ is the Christoffel
symbol and the Dirac matrices in curve space--time $\gamma$ are
connected to the flat space--time Dirac matrices $\bg$ in the
following way
\begin{align}
\gamma_\beta = e_\beta^{(b)} \bg_b, \quad \gamma^\alpha =
e^\alpha_{(a)} \bg^a, \label{gamdir}
\end{align}
where $e^\alpha_{(a)}$ and $e_\beta^{(b)}$ are the tetrad vectors
such that
\begin{align}
g_{\mu\nu} (x) = e_\mu^a(x) e_\nu^b (x) \eta_{ab}, \label{gamtet}
\end{align}
and fulfil the following relations
\begin{align}
e^\alpha_{(a)} e_\beta^{(a)} = \delta_\beta^\alpha, \quad
e^\alpha_{(a)} e_\alpha^{(b)} = \delta_a^b. \label{tetrel}
\end{align}
Here $\eta_{ab} = {\rm diag}(1,\,-1,\,-1,\,-1)$ is the Minkowski
spacetime. The $\gamma$ matrices obey the following anti-commutation
rules
\begin{align}
\gamma_\mu \gamma_\nu + \gamma_\nu \gamma_\mu = 2 g_{\mu\nu}, \quad
\gamma^\mu \gamma^\nu + \gamma^\nu \gamma^\mu = 2 g^{\mu\nu}.
\label{gamcom}
\end{align}

Varying the Lagrangian \eqref{lspin} with respect to $\bp$ and
$\psi$, respectively, we obtain the following spinor field equations
\begin{subequations}
\label{speq}
\begin{align}
\imath\gamma^\mu \nabla_\mu \psi - m \psi - {\cD} \psi - \imath {\cG} \bg^5 \psi & = 0, \label{speq1} \\
\imath \nabla_\mu \bp \gamma^\mu +  m \bp + {\cD} \bp  + \imath
{\cG} \bp \bg^5 & = 0, \label{speq2}
\end{align}
\end{subequations}

where ${\cD} = 2 \lambda  F_K b_1 S, \quad {\cG} = 2 \lambda F_K b_2
P$, \quad $F_K = 	dF/dK$.

The energy-momentum tensor of the spinor field is defined in the
following way \cite{SahaPRD2001}

\begin{align}
T_{\mu}^{\,\,\,\rho}& = \frac{\imath}{4} g^{\rho\nu} \biggl(\bp
\gamma_\mu \nabla_\nu \psi + \bp \gamma_\nu \nabla_\mu \psi -
\nabla_\mu \bar \psi \gamma_\nu \psi - \nabla_\nu \bp \gamma_\mu
\psi \biggr) - \delta_{\mu}^{\rho} L, \label{emt0}
\end{align}
which in view of \eqref{CVD} we rewrite as

\begin{align}
T_{\mu}^{\,\,\,\rho}& = \frac{\imath}{4} g^{\rho\nu} \biggl(\bp
\gamma_\mu
\partial_\nu \psi + \bp \gamma_\nu \partial_\mu \psi -
\partial_\mu \bar \psi \gamma_\nu \psi - \partial_\nu \bp \gamma_\mu
\psi \biggr)\nonumber\\
& - \frac{\imath}{4} g^{\rho\nu} \bp \biggl(\gamma_\mu \Omega_\nu +
\Omega_\nu \gamma_\mu + \gamma_\nu \Omega_\mu + \Omega_\mu
\gamma_\nu\biggr)\psi
 \,- \delta_{\mu}^{\rho} L. \label{temsp01}
\end{align}

Note that the non-diagonal components of the EMT arise thanks to the second term in \eqref{temsp01}. Moreover, let us emphasize that in view of the spinor field equations \eqref{speq}, the spinor field Lagrangian \eqref{lspin} can be expressed as  
\begin{align}
	L = \lambda \left(2 K F_K - F\right). \label{lags1}
\end{align}
We exploit this form of Lagrangian in solving the Einstein equations, as they 
should be consistent with the Dirac one, as \eqref{lags1} is valid only when spinor fields obey the Dirac equations \eqref{speq}. Let us also note that in case $F = \sqrt{K}$ the Lagrangian vanishes, which is very much expected as in this case the spinor field becomes linear.  We are interested in the nonlinear spinor field because only it can generate different kinds of source fields.

\vskip 1 cm

The isotropic and homogeneous cosmological model proposed by
Friedmann, Lemaitre, Robertson, and Walker independently are the most
popular and thought to be realistic, one of the cosmologists. Let
us consider the FLRW model in spherical coordinates in its standard form 
\cite{Narlikar}:
\begin{equation}
ds^2 = dt^2 - a^2(t)\left[\frac{dr^2}{1 - k r^2} + r^2 d\vartheta^2
+ r^2 \sin^2{\vartheta} d\phi^2\right], \label{frwsph}
\end{equation}
with $k$ taking the values $+1,\,0$ and $-1$ which corresponds to a
closed, flat, and open universe, respectively. Though the value of $k$ defines the type 
of geometry of space-time, in reality, it is defined by the contents that fill universe. As we see later, independ to the value of $k$ the universe filled with dark energy is always open, whereas for perfect fluid the value of $k$ really matters. In this case depending on the value of $k$ we obtain close, flat or open universe.

In view of \eqref{gamtet}, the tetrad we will choose in the form

\begin{align}
e^{(0)}_0 = 1, \quad e^{(1)}_1 = \frac{a}{\sqrt{1 - k r^2}},\quad
e^{(2)}_2 = a r, \quad e^{(3)}_3 = a r \sin \vartheta. \nonumber
\end{align}
Then from \eqref{gamdir} we find the following $\gamma$ matrices

\begin{align}
\gamma^0 = \bg^0,\, \quad \gamma^1 =\frac{\sqrt{1 - k r^2}}{a}
\bg^1,\, \quad \gamma^2 = \frac{\bg^2}{a r},\, \quad \gamma^3 =
\frac{\bg^3}{a r \sin \vartheta}. \nonumber
\end{align}

Further from $\gamma_\mu = g_{\mu \nu} \gamma^\nu$ one finds the
$\gamma_\mu$ as well.

The Christoffel symbols, Ricci tensor, and scalar curvature, and the
Einstein tensor corresponding to the metric \eqref{frwsph} is well
known and can be found in \cite{Narlikar}.

Then from \eqref{SPAC} we find the following expressions for spinor
affine connection

\begin{subequations}
\label{SACFRWSph}
\begin{align}
\Omega_0 & = 0, \label{SACFRWSph0}\\
\Omega_1 & = \frac{1}{2\sqrt{1 - kr^2}} \dot a \bg^1 \bg^0, \label{SACFRWSph1}\\
\Omega_2 & = \frac{1}{2} r \dot a \bg^2 \bg^0 + \frac{1}{2} \sqrt{1
- kr^2} \bg^2 \bg^1, \label{SACFRWSph2}\\
\Omega_3 & = \frac{1}{2} \dot a r \sin \vartheta \bg^3 \bg^0 +
\frac{1}{2} \sqrt{ 1 - k r^2} \sin \vartheta \bg^3 \bg^1 +
\frac{1}{2} \cos \vartheta \bg^3 \bg^2. \label{SACFRWSph3}
\end{align}
\end{subequations}

Let us consider the case when the spinor field depends on $t$ only,
then in view of \eqref{SACFRWSph} the spinor field equations can be
written as

\begin{subequations}
\label{speqex}
\begin{align}
\dot \psi + \frac{3}{2} \frac{\dot a}{a} \psi +  \frac{\sqrt{1 - k
r^2}}{ar} \bg^0 \bg^1 \psi + \frac{\cot \vartheta}{2 a r} \bg^0
\bg^2 \psi + \imath \left(m + {\cD}\right) \bg^0 \psi + {\cG} \bg^5 \bg^0 \psi & = 0, \label{speq1n} \\
\dot \bp + \frac{3}{2} \frac{\dot a}{a} \bp - \frac{\sqrt{1 - k
r^2}}{ar} \bp \bg^0 \bg^1 - \frac{\cot \vartheta}{2 a r} \bp \bg^0
\bg^2 - \imath \left(m + {\cD}\right) \bp \bg^0  + {\cG} \bp \bg^5
\bg^0 & = 0. \label{speq2n}
\end{align}
\end{subequations}

The non-trivial components of the energy-momentum tensor of the spinor field in this case read

\begin{subequations}
\label{EMTc}
\begin{align}
T_0^0 & =   m S + \lambda F := \varepsilon, \label{00f} \\
T_1^1 & =  T_2^2 = T_3^ 3 = - \lambda \left( 2 K F_K -
F\right) := -p, \label{iif}\\
T^1_3 & = \frac{a \cos \vartheta}{4\sqrt{1 - kr^2}}\,A^0, \label{13f}\\
T^0_1 & = \frac{\cot \vartheta}{4 r \sqrt{1 - k r^2}} A^3, \label{10f}\\
T^0_2 & = -\frac{3}{4}\sqrt{1 - kr^2}\,A^3, \label{02f}\\
T^0_3 & =  \frac{3}{4} \, \sqrt{1 - kr^2}\, \sin \vartheta A^2 -
\frac{1}{2} \cos \vartheta A^1. \label{03f}
\end{align}
\end{subequations}

From \eqref{EMTc} we conclude that the energy-momentum tensor of the
spinor field contains nontrivial non-diagonal components. It can be seen that the
non-diagonal components do not depend on the spinor field nonlinearity. They 
occur due to the spinor affine connections depending on space-time geometry as well as the system of
coordinates. They impose restrictions on spinor field and/or space-time geometry.

It should be emphasized that for a FLRW model given in Cartesian
coordinates, the EMT has only diagonal components with all the
non-diagonal ones being identically zero \cite{SahaAPSS2020}. So in
this case, the non-diagonal components arise as a result of
the coordinate transformation. Note also that all cosmological space-times defined by diagonal matrices of Bianchi type $VI$, $VI_0$, $V$, $III$, $I$, $LRS-BI$, and $FLRW$ have nontrivial non-diagonal elements that differ from each other in different cases
\cite{2018EChAYa49-146-212}. Moreover, non-diagonal metrics such as
Bianchi type $II$, $VIII$ and $IX$ also have nontrivial non-diagonal
components of EMT. Consequently, we see that the appearance of non-diagonal components of the energy-momentum tensor occurs either due to coordinate transformations or due to the geometry of space-time.

As one sees, the components of the EMT of the spinor field contain
some invariants generated from bilinear spinor forms. From \eqref{speqex} one obtains the following system of equations for the invariants in question:

\begin{subequations}
\label{invariants}
\begin{align}
\dot S_0 + 2 {\cal G} A_0^0 & = 0, \label{S}\\
\dot P_0 - 2 \left( m + {\cal D}\right) A_0^0 & =  0, \label{P}\\
\dot A_0^0 + 2 {\cal G} S_0  + 2 \left( m + {\cal D}\right) P_0 +
2\frac{\sqrt{1 - k r^2}}{ar} A_0^1 + \frac{\cot \vartheta}{ar} A_0^2 & = 0, \label{A0}\\
\dot A_0^1 + 2\frac{\sqrt{1 - k r^2}}{ar} A_0^0 & = 0, \label{A1}\\
\dot A_0^2 + \frac{\cot \vartheta}{ar} A_0^0 & = 0, \label{A2}
\end{align}
\end{subequations}
that gives the following relation between the invariants:

\begin{equation}
P_0^2 - S_0^2 + \left(A_0^0\right)^2 - \left(A_0^1\right)^2 -
\left(A_0^2\right)^2 = C_0, \quad C_0 = {\rm Const.} \label{inv}
\end{equation}

It can be shown that 
\begin{align}
K = \frac{C_0}{a^{6}}, \quad C_0 = {\rm const}. \label{Kdef0}
\end{align}	
Note that \eqref{Kdef0} holds for a massless spinor field in case $K =\{J,\, I\pm J\}$, whereas for $K = I$ it is true for both massive and massless spinor fields. 

Let us recall that the Einstein tensor $G_\mu^\nu$ corresponding to
the metric \eqref{frwsph} possesses only nontrivial diagonal
components. Hence, the general Einstein system of equations
\begin{align}
G_\mu^\nu = - 8 \pi G T_\mu^\nu, \label{GEE}
\end{align}
leads to the following expressions
\begin{align}
0 = T_\mu^\nu, \quad \mu \ne \nu. \label{GEEND}
\end{align}
In view of \eqref{13f} - \eqref{03f} from \eqref{GEEND} one dully
finds that
\begin{align}
A^0 = 0, \quad A^3 = 0, \quad A^1 = (3/2) \sqrt{1 - kr^2}\, \tan
\vartheta A^2. \label{a12}
\end{align}
Note that since the FLRW model is given in Cartesian coordinates, the
non-diagonal components of EMT are identically zero, hence the relation
such as \eqref{a12} does not exist.
In view of $A^0 = 0$, $A^3 = 0$ from the system \eqref{invariants}
we find
\begin{align}
S_0 = C_S, \quad P_0 = C_P, \quad A_0^1 = C_0^1, \quad A_0^2 =
C_0^2, \label{invfrw}
\end{align}
with $C_S$, $C_P$, $C_0^1$ and $C_0^2$ being some arbitrary
constants. Thus we see that $K \propto a^{-6}$. Note that
the equation \eqref{A0} in this case in redundant and \eqref{a12}
gives relations between the constants $C_0^1$ and $C_0^2$.

We are now ready to consider the diagonal components of the Einstein
system of equations which for the metric \eqref{frwsph} takes the
form
\begin{subequations}
\label{ein2}
\begin{align}
2 \frac{\ddot a}{a} +\left( \frac{{\dot a}^2}{a^2} +
\frac{k}{a^2}\right) & = - 8 \pi G p, \label{ein11m}\\
3 \left( \frac{{\dot a}^2}{a^2} + \frac{k}{a^2}\right) & = 8\pi G
\varepsilon. \label{ein00m}
\end{align}
\end{subequations}
On account of \eqref{ein00m} we rewrite \eqref{ein11m} in the form
\begin{equation}
\frac{\ddot a}{a} = 
-\frac{4 \pi G}{3}\left(\varepsilon + 3 p\right), \label{accel}
\end{equation}
where $\varepsilon$ and $p$ are the the energy density
and pressure, respectively defined by \eqref{00f} and \eqref{iif}.
On account of \eqref{00f} and \eqref{iif} from \eqref{accel} we find
\begin{equation}
\ddot a = -\frac{4 \pi G}{3}\left(m S - 2 \lambda F + 6 \lambda K
F_K\right)a. \label{acceln}
\end{equation}

Note that the equations \eqref{accel} or \eqref{acceln} do not
contain $k$ that defines the type of space-time curvature. In order
to take this very important quantity into account, we have to exploit
\eqref{ein00m} as the initial condition for $\dot a$. The equation
\eqref{ein00m} we rewrite in the form
\begin{align}
\dot a = \pm \sqrt{\left(8 \pi/3\right) G \varepsilon a^2 - k} = \pm
\sqrt{\left(8 \pi/3\right) G \left(m S + \lambda F\right) a^2 - k},
\label{vel}
\end{align}
Now we can solve \eqref{acceln} with the initial condition given by
\eqref{vel}. It turns out that these equations are consistent when one
takes the negative sign in \eqref{vel}. Alternatively, one can solve
\eqref{vel}, but for the system to be consistent, he/she has to check
whether the result satisfies \eqref{acceln}.

As we have already established, $S$,\, $K$, hence $F(K)$ are the
functions of $a$. Consequently, given the spinor field nonlinearity the foregoing equation can be solved either analytically or numerically.
The equation \eqref{acceln} can be solved analytically. The first
integral of \eqref{acceln} takes the form
\begin{align}
\dot a = \sqrt{\int f(a) da + C_c}, \label{1stint}
\end{align}
where we define $f(a) =  -\frac{8 \pi G}{3}\left(m S - 2 \lambda F +
6 \lambda K F_K\right)a$ and $C_c$ is a constant which should be
defined from \eqref{vel}. The solution to the equation
\eqref{1stint} can be given in quadrature

\begin{equation}
\int \frac{da}{\sqrt{\int f(a) da + C_c}} = t. \label{quadrature}
\end{equation}

In what follows, we solve the system \eqref{ein2} numerically. In
doing so, we rewrite it in the following way:

\begin{subequations}
\label{ein2a}
\begin{align}
\dot a &= H a, \label{HCsys}\\
\dot H &= -\frac{3}{2} H^2 - \frac{1}{2} \frac{k}{a^2} - 4 \pi G \lambda \left( 2 K F_K - F\right), \label{ein11msys}\\
H^2 & = \frac{8\pi G}{3} \left(m S + \lambda F\right) -
\frac{k}{a^2}, \label{ein00msys}
\end{align}
\end{subequations}
where $H$ is the Hubble constant.

As one sees, in the foregoing system, the first two are differential
equations, whereas the third one is a constraint, which we use as
the initial condition for $H$:
\begin{align}
H  = \pm \sqrt{8\pi G\left(m S + \lambda F\right)/3 - k/a^2}.
\label{Hubin}
\end{align}
Since the expression under the square root must be non-negative, it imposes some restrictions on the choice of the initial value of $a$ as well. Note that the initial value of $H$ depends on spinor mass $m$, coupling parameter $\lambda$, and the value of $k$.

\subsection{Model Construction}
In what follows, we solve the equations \eqref{HCsys} and \eqref{ein11msys} numerically. The third equation of the system \eqref{ein2a} is the initial condition for $H(t)$ in the form \eqref{Hubin}. 
We do it for both massive and massless spinor fields. Besides this, we consider a closed, flat, and open universe, choosing different values for $k$. As it was mentioned earlier, the coupling constant $\lambda$ can be positive or negative. We consider different kinds of spinor field nonlinearities $F(K)$ (equivalently, $F(S)$), that describe various types of sources from perfect fluid to dark energy. 

Since we want to compare our results with observational data, it is convenient to rewrite the system \eqref{ein2a} in terms of red-shift $z$:
\begin{align}
	z &= \frac{\lambda_0 - \lambda_e}{\lambda_e} = \frac{\lambda_0}{\lambda_e} - 1 \longrightarrow  1 + z = \frac{a_0}{a}, \label{red}  
\end{align}
where $\lambda_0$ and $\lambda_e$ are the wavelengths of the photon when detected (now) and emitted, respectively. $a_0$ is the scale factor now and can be taken to be unity: $a_0 = 1$. 
 Then from \eqref{Kdef0} we find  
\begin{align}
K &= \frac{K_0}{a^6} = K_0 (1 + z)^6, \quad K_0 = {\rm const.} \label{Kdef} 	
\end{align}

Then we can write the foregoing system as follows: 

\begin{subequations}
	\label{ein2az}
	\begin{align}
		\frac{\partial a}{\partial z} &= - \frac{a_0}{(1+z)^2}, \label{HCsysz}\\
		\frac{\partial H}{\partial z}&= \frac{1}{1+z}\left[ H + \frac{\kappa}{6H} \left( \varepsilon + 3 p \right)\right], \label{ein11msysz}\\
		\frac{\partial \varepsilon}{\partial z} &= \frac{3}{1+z} \left(\varepsilon + p\right).\label{Conservsysz} 
	\end{align}
\end{subequations}

To solve the equation \eqref{ein2az} numerically, we have to give the concrete form of the spinor field nonlinearity. It was found earlier that the spinor field nonlinearity can simulate different types of source fields, such as quintessence, Chaplygin gas, modified quintessence, modified Chaplygin gas, etc. Since in all cases the Lagrangian becomes massless, we consider only a massless spinor field, whereas the nonlinear term will play the role of effective mass. Thus, setting $\varepsilon = \lambda F(K)$ and $p = \lambda \left(2 K F_K - F(K)\right)$, we find the following expression for the modified Chapligyn gas as a spinor field DE model given by the equation of state:
\begin{align}
	p = W \ve - A/\ve^\alpha, \label{mchap}
\end{align}
with $W$ being a constant, $A > 0$ and $0 \le \alpha \le 1$. 
Though the dark energy and the dark matter act in a completely
different way, many researchers suppose that they are different
manifestations of a single entity. In fact, if $W$ is taken to be positive, then in initial stage
of evolution energy density was quite large, hence the first term prevails and pressure becomes positive, while with the expansion of the universe energy density becomes very small and the second term predominates giving rise to a negative pressure that conforms to the dark energy. Following such an idea, a modified Chaplygin gas was introduced in \cite{B7} and was further developed
in \cite{Benaoum2012}.

Inserting $\varepsilon = T_0^0$ and $ p = - T_1^1$ from \eqref{EMTc} into \eqref{mchap} we find the nonlinear term that describes a modified Chapligyn gas:
\begin{align}
	F(K) &= \left[\frac{A}{1+W} + \lambda_1
	K^{(1+\alpha)(1+W)/2}\right]^{1/(1+\alpha)}. \label{modchap0F}
\end{align} 
In the above expressions $\lambda_1$ is the constant of integration. On account of \eqref{Kdef}  energy density and pressure of the system, we rewrite in terms of red-shift: 

 \begin{subequations}
	\begin{align}
		\varepsilon &=  \lambda \left[\frac{A}{1+W} + \lambda_1
		(1+z)^{3(1+\alpha)(1+W)}\right]^{1/(1+\alpha)} ,  \label{mchapsped}\\
		p &=  \lambda \left[\lambda_1 W
		(1+z)^{3(1+\alpha)(1+W)} - \frac{A}{1+W}\right] \left[\frac{A}{1+W} + \lambda_1
		(1+z)^{3(1+\alpha)(1+W)}\right]^{-\alpha/(1+\alpha)}
		. \label{modchapp} 
	\end{align}
\end{subequations}
 The deceleration parameter $q(z)=-\frac{\ddot a}{a}\frac{1}{H^{2}}$ for this MCG model can be constructed by using the equation \eqref{accel}, \eqref{ein00msys}, \eqref{00f}, and \eqref{red}. Hence, one can get for the open universe ($k=-1$) as
 \begin{align}
     q(z)=\frac{(\varepsilon + 3p)/3}{\varepsilon /3 +(z+1)^{2}}. \label{Decpar}
 \end{align}
 As one sees, in the case of dark energy, when $\varepsilon + 3p < 0$, which corresponds to the accelerated expansion.  

\begin{figure}[h!]
\centering
\includegraphics[width=0.8\textwidth]{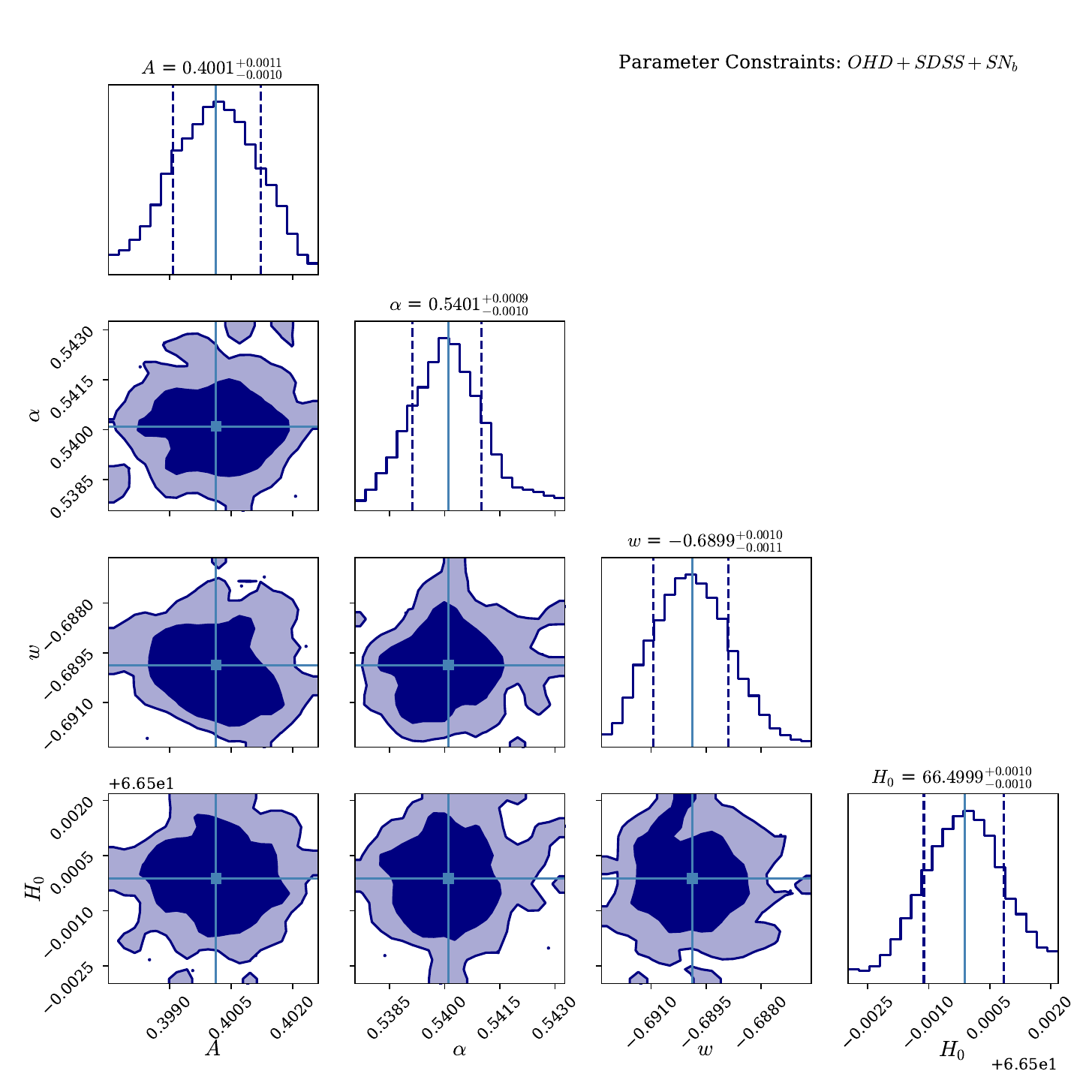}
\caption{Marginalized posterior distributions of the MCG model parameters for Combined OHD + SDSS + SN (binned).}
\end{figure}

\begin{figure}[h!]
\centering
\includegraphics[width=0.8\textwidth]{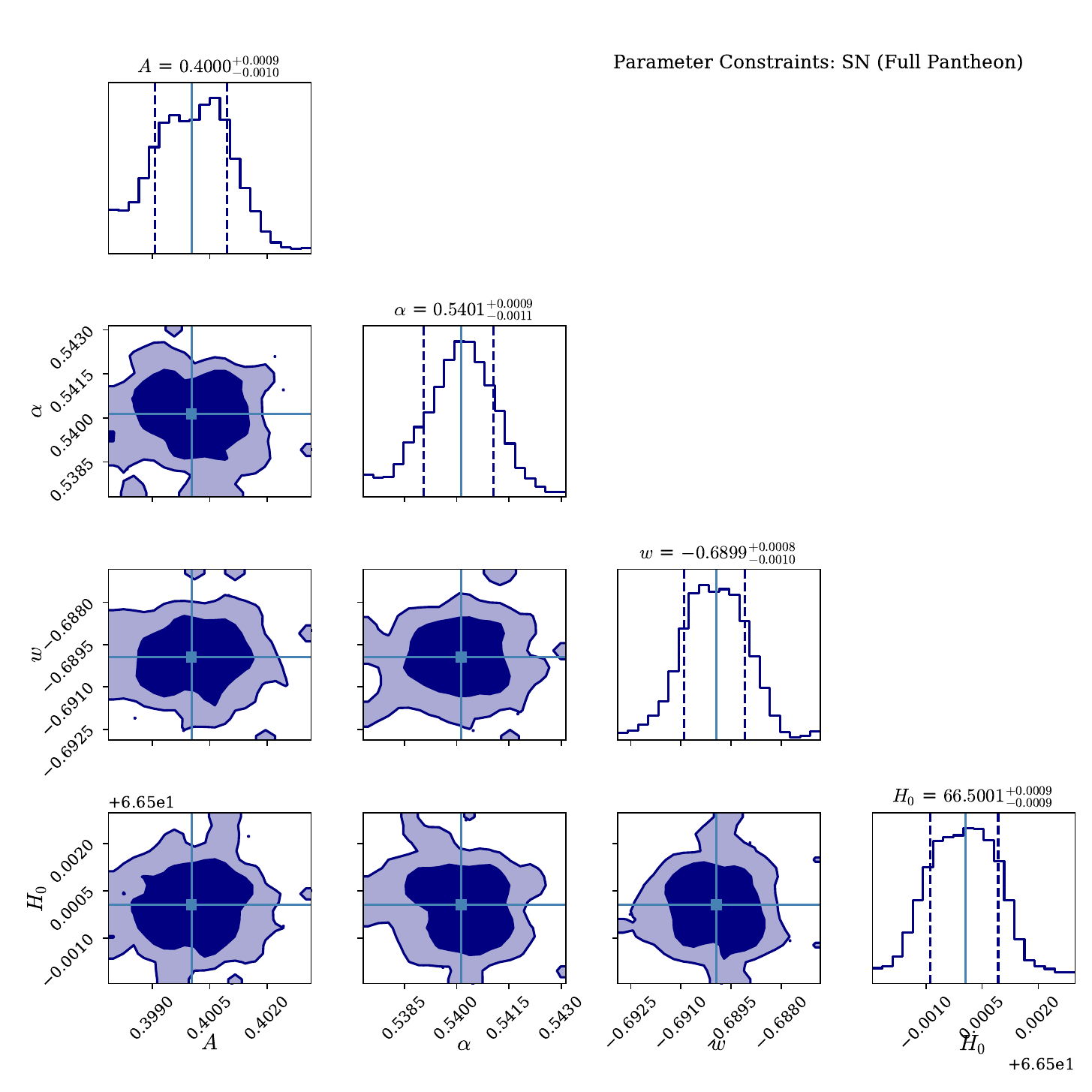}
\caption{Marginalized posterior distributions of the model parameters for the Full Pantheon data set.}
\end{figure}

\section{Observational data and Methodology}
To parametrize the  MCG model in the scenario of spherically symmetric FLRW spinor field, we considered 43 OHD Hubble measurements from CC \cite{Stern2010, Moresco2012, Moresco2016, Zhang2014, Simon2005}, and 48 SDSS $H(z)$ data points from BAO measurements \cite{Delubac2015, Alam2017}. Additionally, 11 Pantheon (binned) points with errors from SNIa \cite{Scolnic2018}. Also, the full Pantheon 1078 data set has been simulated separately. All the data sets are spread within the range of $0<z<2.5$ in the different redshifts. To constrain the model parameters, we used \texttt{emcee} Python library for performing the MCMC technique in Python 3 ($\it{ipykernel}$). For the MCMC simulation, the likelihood $\chi^{2}$ can be obtained as 
\begin{align}
  \chi^{2}= \chi^{2}_{OHD}+\chi^{2}_{SDSS}+\chi^{2}_{SNb}
\end{align}
The $\chi^{2}_{OHD},~\chi^{2}_{SDSS}$, and $\chi^{2}_{SNb}$ correspond to OHD, SDSS, and binned Supernova (SNb) data sets, respectively. The $\chi^{2}$ statistic is given by:
\begin{equation}
\chi^2 = \sum_{i=1}^{N} \left( \frac{O_i - T_i}{\sigma_i} \right)^2
\end{equation}
where: \( O_i \), \( T_i \), \( \sigma_i \), and \( N \) are the observed value, theoretical (model) value,  uncertainty, and total number of data points, respectively.

We use to explore the four-dimensional parameter space; paramas~$ =A, \alpha, W, H$ where $A$, $\alpha$ are model parameters. $W$ stands for EoS, and $H$ represents the Hubble parameter. Several observations esteemed the present value of the Hubble parameter lies between $H_{0}~(67.0-74.03)$ km/s/Mpc \cite{Planck2018, Riess2021, Freedman2021, Verde2019, DiValentino2021}. However, in our analysis, we took the priors of the Hubble parameter in $50 < H< 90$, and other priors to the model parameters: $0 < A < 2$, $0 < \alpha < 1.5$, and $-1 < W < 1$. For MCMC simulation, we set the value of others two constants of the MCG model as $\lambda =0.98$ and $\lambda_{1}=1.0$. To run the MCMC, we set walkers $=200$ and steps $=5000$. 

\begin{figure}[h!]
\centering
\includegraphics[width=0.47\textwidth]{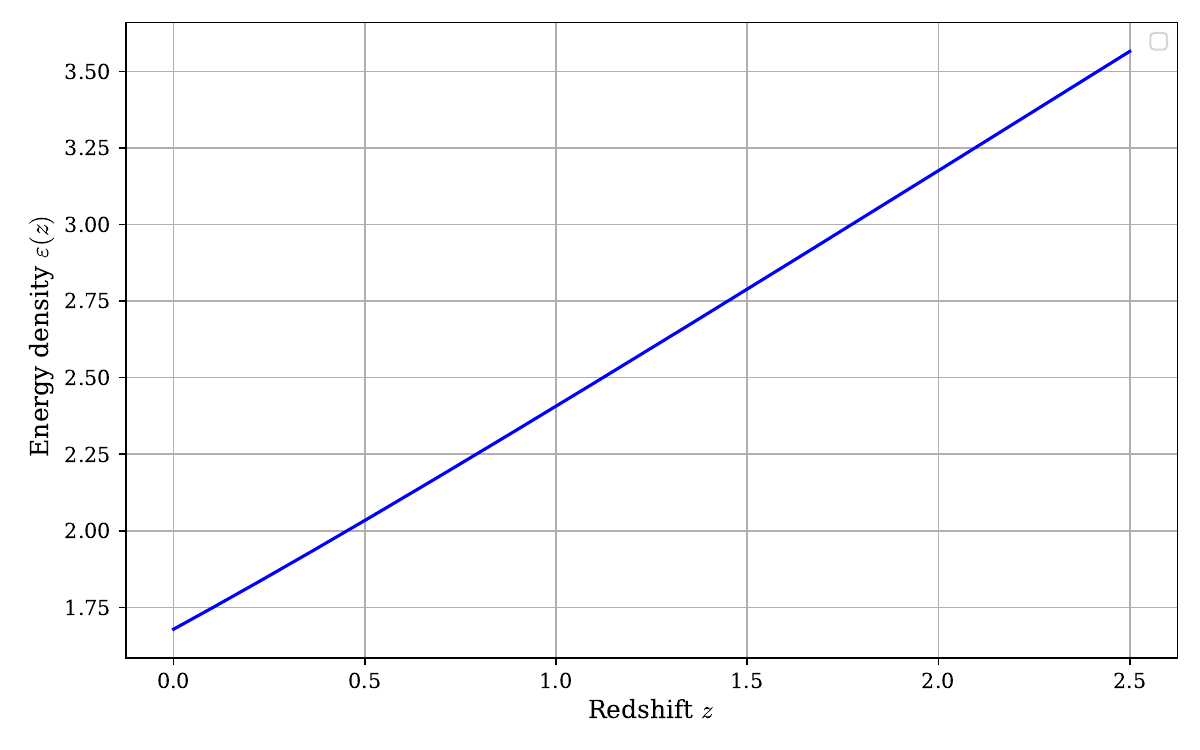}
\includegraphics[width=0.47\textwidth]{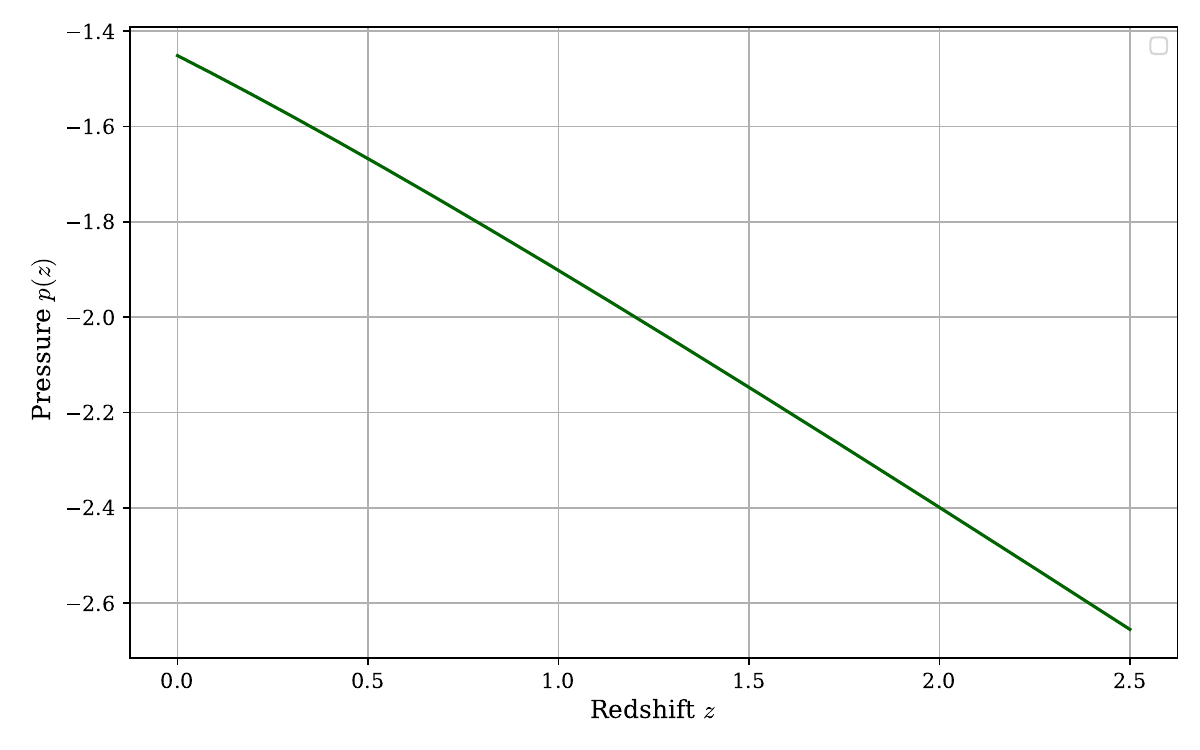}
\caption{Evolution of energy density $\varepsilon(z)$ (Left), and pressure $p(z)$ (Right) in the late-time universe for spinor field DE model.}
\end{figure}

\begin{figure}[h!]
\centering
\includegraphics[width=0.8\textwidth]{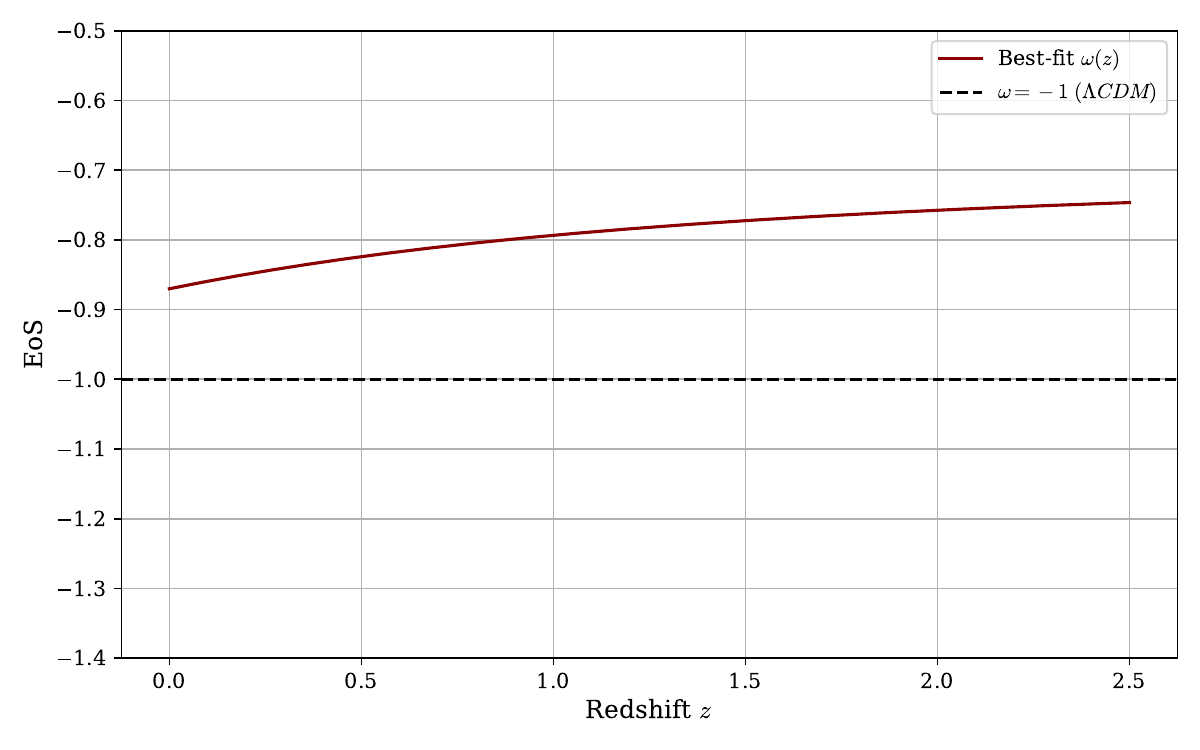}
\caption{Evolution of the equation of state $w(z)$ for the spinor field DE model.}
\end{figure}

\section{Numerical results}
The numerical results of the MCMC simulation on the MCG model are summarized in Table 1.  The results come from the sampling of two different sets of observational data. Sampling of the combined data set $OHD+SDSS+SNb$ and $SN$ (Full Pantheon ) produces almost the same results. In Table 1, Columns $III$ and $IV$ show the best-fit results of the parameters with error for the data set  $OHD+SDSS+SNb$ and $SN$ (Full Pantheon ), respectively, and in Column $II$ we have shown the priors of the parameters. The normalized best-fit value of parameters from MCMC analysis is $[A, \alpha, W, H_0: 0.4000, 0.5401, -0.6899, 66.5000]$. The best-fit value of our model's parameters closely aligns with the other MCG models. To see a clear visual of the posterior distribution of the MCG model, we may look at Figure 1 and Figure 2, which are marginalized one-dimensional and two-dimensional plots, with a $1\sigma$ confidence level ($68\%$) for the combined data set $OHD+SDSS+SNb$ and $SN$ (Full Pantheon ) respectively. These two contour plots fully correspond to Table 1. The reduced chi-square statistical measurement of Hubble data points $\chi^2_{\mathrm{red}} = 1.08$, which indicates that the model is a good fit with the observational data.

\begin{table}[ht]
\centering
\caption{Best fit values of different cosmological parameters of MCG model for different combinations of data sets.}
\vspace{0.1 cm}
\begin{tabular}{lccc}
\hline
\textbf{Parameter} & \textbf{Priors} &\textbf{OHD + SDSS + SNb} & \textbf{ SN(Full Pantheon)} \\
\hline
\vspace{0.1 cm}
$A$          &[0, 2.0] & $0.4001^{+0.0011}_{-0.0010}$  & $0.4000^{+0.0009}_{-0.0010}$ \\
\vspace{0.1 cm}
$\alpha$     & [0, 1.5] &$0.5401^{+0.0009}_{-0.0010}$  & $0.5401^{+0.0009}_{-0.0010}$ \\
\vspace{0.1 cm}
$W$          & [-1, 1] &$-0.6899^{+0.0010}_{-0.0011}$  & $-0.6899^{+0.0008}_{-0.0010}$ \\
\vspace{0.1 cm}
$H_0$ [km/s/Mpc] & [50, 90] &$66.4999^{+0.0010}_{-0.0010}$ & $66.5001^{+0.0009}_{-0.0009}$ \\
\hline
\end{tabular}
\label{tab:mcg_constraints}
\end{table}

To understand the accelerated expansion of the universe for the spinor field DE model, we plot the changing pressure and energy density with redshift (Figure 3). On the right side of the figure, we can see that the pressure is positive at high redshift, and becomes negative at lower redshift (toward $z\xrightarrow {}0$). Similarly, the evolution of energy density $\varepsilon (z)$ (Left-side of Figure 3) indicates that the Spinor-DE model clearly describes an accelerating universe at low redshift with the rise of $\varepsilon (z)$ toward higher $z$. 

In Figure 4, this Spinor field model behaves as an alternative to the $\Lambda CDM$ model, EoS is evolving very smoothly across the redshift ($w(0) = -0.8701, w(2.5) = -0.7464$), unlike the $\Lambda CDM (\omega(z)=-1)$. The EoS of our spinor field MCG model is $\omega >-1$ at $z\xrightarrow {}0$ like other MCG models. The present value of the deceleration parameter $q(z)$ is another validation of this DE model. The best-fit deceleration parameter would typically relate to the expansion of the universe. In a late-time universe experiencing accelerated expansion, the deceleration parameter $q(z)$ is negative. In Figure 5, we may see the recent $q(z=0)=-0.571$. Here is a comparison with other DE models $[\Lambda CDM(\omega=-1), wCDM(\omega \approx -0.9), Quintessence~(\omega > -1), Phantom~(\omega < -1): -0.55, -0.5~to -0.6, -0.5~to -0.6, -0.6~to -0.7]$. Hence, our result of $q(z)$ is pretty much aligned with the other models, that indicates the model is best fit for the late-time expansion of the universe. 

We can also notice the evolution of the Hubble parameter with redshift in Figure 6. The best-fit line is closely aligned with observational OHD and SDSS data sets; the black solid line represents our Spinor-field DE model and compares with the $\Lambda CDM$ model(black dashed line). 

\begin{figure}[h!]
\centering
\includegraphics[width=0.8\textwidth]{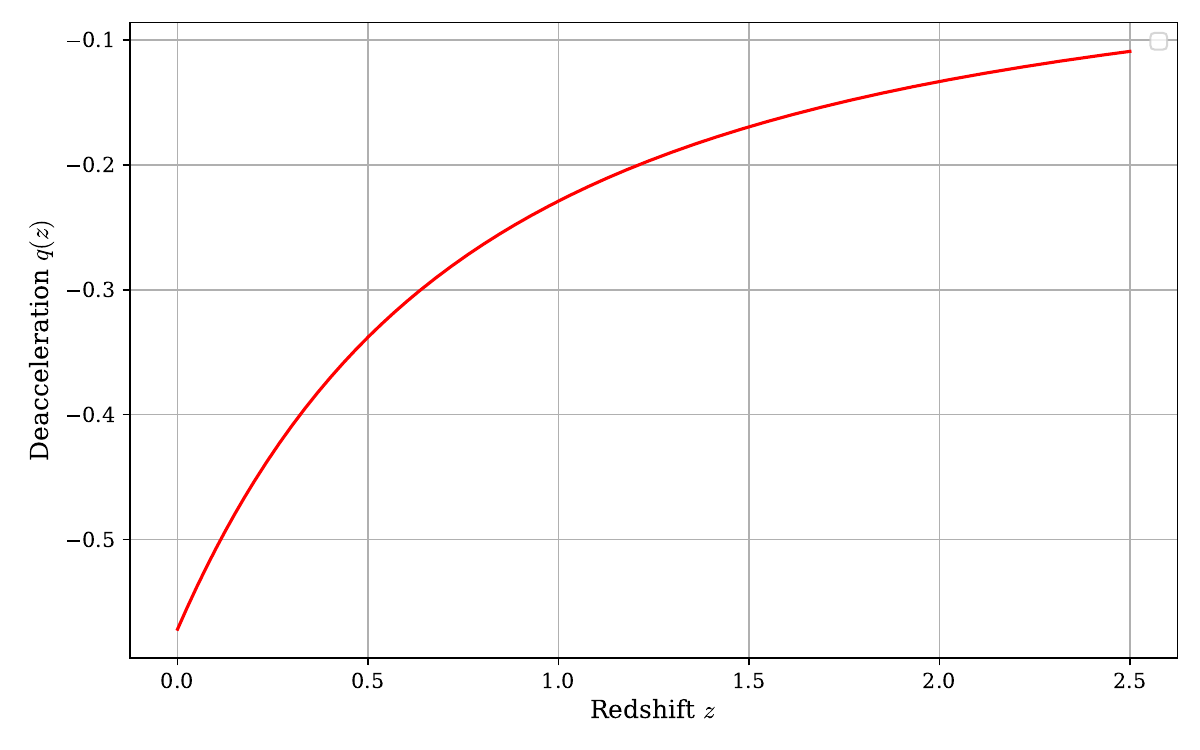}
\caption{Best fit deceleration parameter $q(z)$ with redshift $(z)$ in the late-time universe indicating the acceleration expansion.}
\end{figure}

\begin{figure}[h!]
\centering
\includegraphics[width=0.8\textwidth]{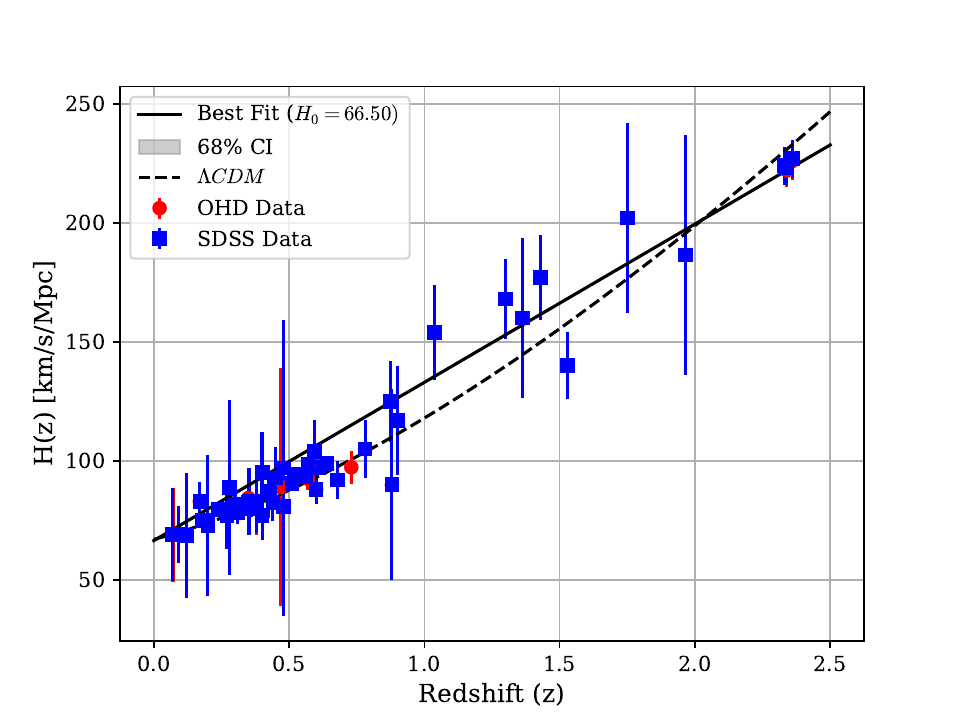}
\caption{Comparison of best-fit $H(z)$ of the spinor field MCG model to OHD and SDSS data along with $\Lambda CDM$ model.}
\end{figure}

In Figure 7(Right), we compare the model of the best-fit distance modulus $(\mu_{bestfit})$ with binned supernova data from distant galaxies. The best-fit $\mu_{bestfit}$ is moderately aligned with the data points. Also, we calculated the residual of distance modulus, shown in Figure 7(Right), where the residual is given as $\Delta {\mu}=\mu_{observation}-\mu_{bestfit}$. Most of the residual points belong to $-1~to +1$.

\begin{figure}[h!]
\centering
\includegraphics[width=0.47\textwidth]{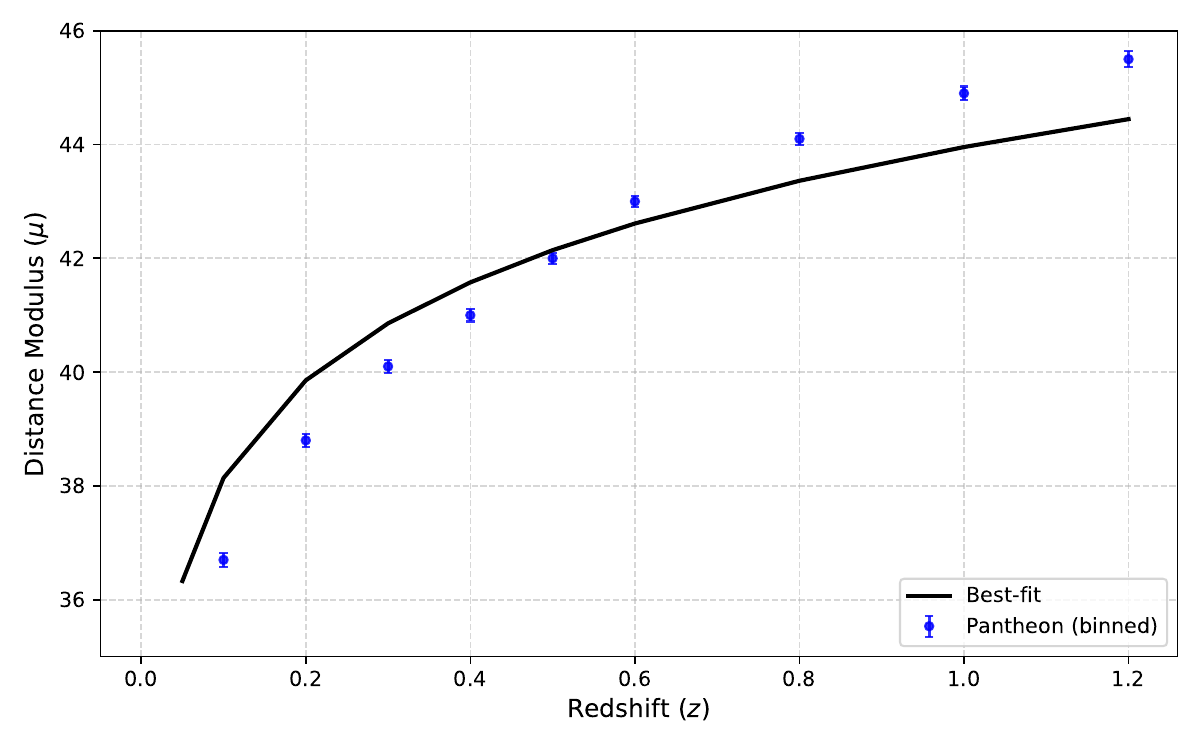}
\includegraphics[width=0.47\textwidth]{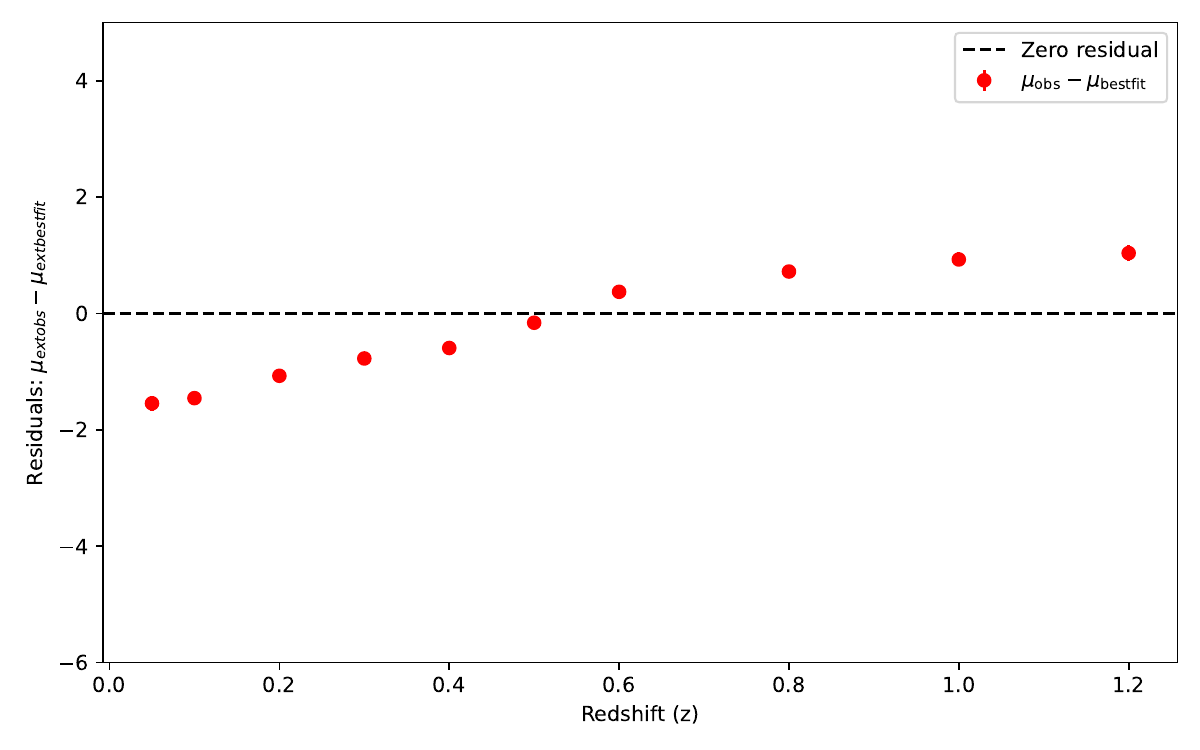}
\caption{Left: Best-fit $\mu(z)$ curve compared with Pantheon binned data. Right: Residuals; $\mu_\text{obs} - \mu_\text{bestfit}$.}
\end{figure}

\section{Conclusion}
In the context of a spherically symmetric FLRW model, we found that the spinor field exhibits nontrivial non-diagonal components in its energy-momentum tensor (EMT). Since the Einstein tensor in this framework is diagonal, this leads to specific constraints on the spinor field components:  $A^0 = 0$, $A^3 = 0$, and $A^1 \propto A^2$. The corresponding field equations were solved under these conditions. Our results show that when the spinor field nonlinearity mimics ordinary matter, such as radiation, the curvature parameter 
$k$ becomes crucial—yielding a closed, flat, or open universe depending on whether $k$ is positive, zero, or negative, respectively. Additionally, the spinor mass affects the outcome quantitatively. However, when the spinor field nonlinearity models dark energy, the universe undergoes rapid expansion regardless of the value of $k$.

It should be noted that the spinor description of the source field introduces a number of unexpected features. The presence of non-diagonal components in the energy-momentum tensor imposes restrictions not only on the space-time geometry but also on the spinor field itself. On the other hand, since the law of energy conservation is automatically satisfied due to the spinor field equations, one can readily consider a variety of source types — ranging from ordinary matter to dark energy — which may correspond to different stages of cosmic evolution.

In this article, we have studied the role of a nonlinear spinor field in the universe's evolution under the spherically symmetric FLRW space-time. This model offers a viable alternative to $\Lambda$CDM in explaining late-time cosmic acceleration, fitting SN, OHD, and SDSS data with competitive accuracy. The best fit value of $H_{0}$ as determined from numerical analysis is $66.4999^{+0.0010}_{-0.0008}$. This result is compatible with the PLANCK collaboration results. We reconstructed EoS $w(z)$ shows evolution consistent with a MCG-like behavior, and the deceleration parameter $q(z)$ confirms the universe's acceleration in the present epoch. Further, the reduced chi-square statistics of Hubble data points, and the residual of the distance modulus indicate the goodness of the fit of the model in the DE-dominated era.
\\
\\
{\bf Acknowledgement:}\,\,\,\,{We thank Dhiraj Hazra for helpful discussions on MCMC data constraints.}{This paper has been
supported by the RUDN University Strategic Academic Leadership
Program.} 
{}
\\
\\
{\bf Declarations:}

\vskip 5 mm {\bf Competing interests} { There is
no conflict of interests.}


\vskip 5 mm {\bf Funding}  Not applicable.

\vskip 5 mm {\bf Availability of data and materials} No new data sets were generated during the current study.

\end{document}